\documentclass{elsart}

\usepackage{amssymb}
\usepackage{amsfonts}
\usepackage{amsmath}
\usepackage{graphicx}
\usepackage{booktabs}
\usepackage{amssymb}

\begin{document}

\begin{frontmatter}



\title{Theoretical study on RbCl crystal with M\"{o}bius inverse potentials}


\author{Yue Shen, Shuo Zhang, and Nan-xian Chen}

\address{Department of Physics, Tsinghua University, Beijing,
100084, China}

\begin{abstract}
The alkali halides have been studied very frequently for their
simple structures and interesting properties, for example, the
high-pressure induced transition$^1$. There are several successful
models that can be employed. The most famous model of them is the
Tosi-Fumi interionic potential$^2$ which is an empirical potential
derived from the experimental data. Whereas, here we developed a
new potential model based on the M\"{o}bius lattice inversion
method$^{3,4}$, which can be derived directly from the cohesive
energy curve without any experimental data and is more effective
than the ab initial calculation. With the M\"{o}bius interionic
potentials, we calculated the structural and elastic properties of
RbCl crystal. The results are in good agreement with experiments.
We also studied the high-pressure induced B1-B2 transition of RbCl
crystal and estimated approximately the transition-point which is
about 1.09GPa. Further more, we used this potential model to
simulate the RbCl melting with molecular dynamics. The calculated
melt-point is approximately 990K$\sim$995K, close to the
experimental data.
\end{abstract}

\end{frontmatter}

\section{Introduction}
During the past 60 years the alkali halides have been well studied
for their thermodynamic, elastic and structural properties. The
Tosi-Fumi potential model$^2$ is usually employed. Among those
available models that can correctly describe the alkali halides,
the pairwise interionic potentials always include two parts: the
long-range Coulomb term and the short-range term. The Coulomb term
is of the form $Z_iZ_j/r_{ij}$, where $r_{ij}$ represents the
distance between the $i$th and the $j$th ions, $Z_i$ and $Z_j$
represent their effective charges. While there is not a definitive
form of the short-range term. Tosi-Fumi model uses the forms as
$$-\frac{C_{ij}}{r_{ij}^6}-\frac{D_{ij}}{r_{ij}^8}+A_{ij}exp(-B_{ij}r_{ij})$$,
which represent dipole-dipole, dipole-quadrupole, and repulsion
terms respectively. However, if we have the short-range potential
curves, any compatible function forms will be fitted to the
potential curves with adaptive parameters. In pervious models$^2$
the parameterization of potentials is usually obtained by
empirical fitting which depends on experimental data. These
empirical potentials are only applicable with certainty over the
range of interionic distances used in the fitting procedure, which
may lead to problems if the potential is used in a calculation
that accesses distances outside this range. Here we find the way
based on the M\"{o}bius lattice inversion method to avoid these
problems. Using the lattice inversion method we get the
short-range potential curves solely from the cohesive energy
curves which can be calculated with quantum mechanics. It is to
say that our short-range potential curves are derived without any
experimental data. This method will be discussed in detail later.
Another question is how to define the effective charges of cations
and anions. One possible choice is to assign cations and anions
with formal charges as Rb+1 and Cl-1. But here we use another way
to decide the effective charges which is depicted in part II. We
consider nonintegral charges are more reasonable. This paper is
organized as follows: First, we introduce our M\"{o}bius inverse
potential model in detail. Further, some structural and elastic
properties of B1-RbCl crystal are calculated and compared with
experimental data. Second, we study the high-pressure-induced
transition from B1 structure to B2 structure with our model, the
transition point is also provid8ed. Finally, we simulate the
melting of B1-RbCl crystal with molecular dynamics employing our
model and calculate the correct melting temperature with a
coexist-phase method. Thus we come to the conclusion that our
M\"{o}bius inverse potential model is successful in describing the
RbCl alkali halide and this lattice inversion technique should
have a wider perspective.

\section{M\"{o}bius inverse pairwise interionic potentials}
The main idea of M\"{o}bius lattice inversion method can be
described simply as follows$^{5,6}$: For a single element crystal,
the cohesive energy can be expressed as
\begin{equation}
E(x)=\frac{1}{2}\sum_{n=1}^{\infty}r_0(n)\phi(b_0(n)x)
\end{equation}
where $x$ is the nearest-neighbor distance, $r_0(n)$ is the $n$th
neighbor, and $\phi(x)$ is the pair potential. By a
self-multiplicative process from $\{b_0(n)\}$, the $\{b(n)\}$ is
formed, a multiplicative closed semi-group. This implies that a
lot of virtual lattice points are involved, but the corresponding
virtual coordination number is zero. In the $\{b(n)\}$, for any
two integers m and n, there is a sole integer $k$ such that
$b(k)=b(m)b(n)$. Hence, the equation above can be rewritten as
\begin{equation}
E(x)=\frac{1}{2}\sum_{n=1}^{\infty}r(n)\phi(b(n)x)
\end{equation}
where
\begin{equation}
r(n)=\{
\begin{array}{c c}
r_0(b_0^{-1}[b(n)]), & b(n)\in\{b_0(n)\}\\
0, & b(n)\not\in \{b_0(n)\}
\end{array}
\end{equation}
Then the general equation for the pairwise interatomic potential
obtained from M\"{o}bius inversion can be expressed as
\begin{equation}
\phi(x)=2\sum_{n=1}^{\infty}I(n)E(b(n)x)
\end{equation}
where I(n) has the characteristics of
\begin{equation}
\sum_{b(d)|b(n)}I(d)r\bigg(b^{-1}\bigg[\frac{b(n)}{b(d)}\bigg]\bigg)=\delta_{n1}
\end{equation}
Thus we get the pair potential just from the cohesive energy
curve. But we encounter some obstacles when we try to get the
interionic pair potentials from the B1-RbCl cohesive energy curve.
First, we have to know the effective charges of cations and anions
so that we can remove the Coulomb energy from the total energy and
remain only the short-range terms. Then we use M\"{o}bius lattice
inversion to obtain the short-range potentials from the so called
short-range-energy curve. However, there are still three types of
short-range potentials as Rb-Rb, Rb-Cl, Cl-Cl, and it is
impossible to find a lattice constructed only with one kind of
ion. To solve this problem we build several virtual structures of
lattice that may be not real for RbCl crystal. These virtual
structures have some similar traits with the real B1-strucure. For
example, one of them, the B3 structure has the same sublattice of
Rb-Rb and Cl-Cl with the B1 structure. That means the B3 structure
contains the same contributions of cation-cation interaction and
anion-anion interaction with the B1 structure. When the
short-range-energy curve of the B1-structure is subtracted by that
of the B3-structure, we clearly get the energy that contains only
the contribution of cation-anion interaction and its relationship
with lattice constant.

The total energies of B1, B3, T1, and B2 structures are calculated
with CASTEP$^{7,8,9}$ (Cambridge Serial Total Energy Package). The
ultra-soft pseudopotentials for rubidium and chlorine ions are
adopted and the GGA-PW method is used to cope the
exchange-correlation energy. And the k-mesh points over Brillouin
zone are generated with parameters $4\times4\times4$ for the
biggest reciprocal space and $1\times1\times1$ for the smallest
one by the Monkhorst-Pack-scheme$^{10}$ corresponding to lattice
constant a. The energy tolerance for SCF convergence is
$2\times10гн6$ eV/atom, and the kinetic energy cutoff for plane
wave basis set is 260 eV.

To decide the effective charges we process as follows: Since the
short-range parts are quickly convergent when the lattice constant
increases, the total contribution of energy with the lattice
constant larger than 10.0 $\AA$ is almost completely from the
Coulomb part. Then, by using the Madelung constants of B1 and
B3-structure$^{11}$ we can determine the effective charges of ions
from the energy difference between the total energies of B1 and
B3-structure RbCl crystals of large lattice constant.

After having known the effective charges we may calculate the
Coulomb energies of different structures employing the Ewald
summation$^{12}$ and remove this part of energy from the total
cohesive energy. Then we start the lattice inversion from the
short-range-energy curve.

For B1-structure
\begin{equation}
E_{SR}^{B_1}(a)=E_{++}^{B_1}(a)+E_{--}^{B_1}(a)+E_{+-}^{B_1}(a)+E_i
\end{equation}
where ESR means the short-range parts of the total energy,
$E_{++}$, $E_{--}$ and $E_{+-}$ are the energy contributions of
cation-cation, anion-anion and cation-anion respectively, $E_i$
refers to the energy of an isolated ion, and a is the lattice
constant. All energy terms are averaged to each ion. Further we
have
\begin{equation}
E_{++}^{B_1}(a)=\frac{1}{4}\sum_{i,j,k\neq0}\phi_{++}\bigg(\frac{a}{2}\sqrt{(i+k)^2+(i+j)^2+(j+k)^2}\bigg)
\end{equation}
\begin{equation}
E_{--}^{B_1}(a)=\frac{1}{4}\sum_{i,j,k\neq0}\phi_{--}\bigg(\frac{a}{2}\sqrt{(i+k)^2+(i+j)^2+(j+k)^2}\bigg)
\end{equation}
\begin{equation}
E_{+-}^{B_1}(a)=\frac{1}{2}\sum_{i,j,k\neq0}\phi_{+-}\bigg(\frac{a}{2}\sqrt{(i+k-1)^2+(i+j-1)^2+(j+k-1)^2}\bigg)
\end{equation}
Where $\phi_{++}$, $\phi_{--}$, and $\phi_{+-}$ are the
short-range interionic potentials of cation-cation, anion-anion
and cation-anion respectively.

For B3-structure
\begin{equation}
E_{SR}^{B_3}(a)=E_{++}^{B_3}(a)+E_{--}^{B_3}(a)+E_{+-}^{B_3}(a)+E_i
\end{equation}
and
\begin{equation}
E_{++}^{B_3}(a)=\frac{1}{4}\sum_{i,j,k\neq0}\phi_{++}\bigg(\frac{a}{2}\sqrt{(i+k)^2+(i+j)^2+(j+k)^2}\bigg)
\end{equation}
\begin{equation}
E_{--}^{B_3}(a)=\frac{1}{4}\sum_{i,j,k\neq0}\phi_{--}\bigg(\frac{a}{2}\sqrt{(i+k)^2+(i+j)^2+(j+k)^2}\bigg)
\end{equation}
\begin{equation}
E_{+-}^{B_3}(a)=\frac{1}{2}\sum_{i,j,k\neq0}\phi_{+-}\bigg(\frac{a}{2}\sqrt{(i+k-\frac{1}{2})^2+(i+j-\frac{1}{2})^2+(j+k-\frac{1}{2})^2}\bigg)
\end{equation}
As we can see, the difference is just between the cation-anion
interactions.

So we have
\begin{equation}
\begin{split}
\Delta E_{+-}^{SR}(a)&=E_{SR}^{B_1}(a)-E_{SR}^{B_3}(a)\\
&=E_{+-}^{B_1}(a)-E_{+-}^{B_3}(a)\\
&=\frac{1}{2}\sum_{i,j,k\neq0}\phi_{+-}\bigg(\frac{a}{2}\sqrt{(i+k-1)^2+(i+j-1)^2+(j+k-1)^2}\bigg)\\
&-\frac{1}{2}\sum_{i,j,k\neq0}\phi_{+-}\bigg(\frac{a}{2}\sqrt{(i+k-\frac{1}{2})^2+(i+j-\frac{1}{2})^2
+(j+k-\frac{1}{2})^2}\bigg)
\end{split}
\end{equation}
Note that the isolated-ion energy is also gone.

This equation may be rewritten as
\begin{equation}
\Delta
E_{+-}^{SR}(x)=\frac{1}{2}\sum_{n=1}^{\infty}R_{+-}(n)\Phi_{+-}[B_{+-}(n)x]
\end{equation}
in which we substitute lattice constant a with the nearest
cation-anion distance x. Then we follow the way discussed at the
beginning of this part to obtain the curve of $\Phi_{+-}(x)$. The
derivation of cation-cation and anion-anion short-range potentials
is of the same manner. For instance, in order to extract the
anion-anion short-range interaction we build a T1-structure
lattice. The T1-structure lattice has the same sublattice of Rb-Rb
with B1-structure. So the short-range energy difference between
those two structures is dedicated by the cation-anion and the
anion-anion interactions and we can exclude the cation-anion part
since the cation-anion interionic potential is already known. For
the cation-cation one we simply use the B2 and B1-structure
lattices, extract the cation-cation interaction with the
cation-anion and the anion-anion potentials available. The whole
process can be illustrated in Fig.1. After all the
short-range-potential curves are attained, we use several forms of
functions to fit these curves and get the suited parameters. We
choose an exponential repulsive function for cation-anion and a
Morse-stretch function for anion-anion. We find that among the
three short-range potentials the cation-anion one is excessively
more important than the other two while the cation-cation one is
much smaller even than the anion-anion's, almost can be neglected.
This leads us to ignore the cation-cation short-range interaction
for saving calculation time.

So the potentials can be expressed as
\begin{equation}
\Phi_{+-}(x)=D_{+-}exp\bigg[\gamma_{+-}\bigg(1-\frac{x}{R_{+-}}\bigg)\bigg]+
\frac{q_{+}q_{-}}{4\pi\epsilon_0x}
\end{equation}
\begin{equation}
\Phi_{--}(x)=D_{--}\bigg(\bigg\{1-exp\bigg[\gamma_{--}\bigg(1-\frac{x}{R_{--}}\bigg)\bigg]
\bigg\}^2-1\bigg)+\frac{q_{-}q_{-}}{4\pi\epsilon_0x}
\end{equation}
\begin{equation}
\Phi_{++}(x)=\frac{q_{+}q_{+}}{4\pi\epsilon_0x}
\end{equation}
The potential parameters for RbCl are listed in Table I.

With this potential model we calculate some structural and elastic
properties of B1-RbCl crystal at 0K. The results are listed in
Table II. Compared with experimental data at room
temperature$^{13}$, the results show a good agreement.

\begin{table}
\begin{tabular}{ccccccccc}
\toprule &Rb-Cl &&&Cl-Cl &&Effective&
 Charges\\
\midrule $D_{+-}(ev)$ &$R_{+-}(\AA)$ & $\gamma_{+-}$ & $D_{--}(ev)$ &$R_{--}(\AA)$ & $\gamma_{--}$ & $q_{+}$&$q_{-}$\\
\midrule 1.8140 &  2.4470 &     6.7525 &     0.1508 & 3.8632 &    8.5640 & 0.96930e & -0.96930e \\
\bottomrule
\end{tabular}
\caption{Parameters of potential functions}
\end{table}

\section{High-pressure-induced B1-B2 transition}
Most alkali halides crystallize in the B1 (NaCl-like) structure
under ambient temperature and pressure but turn into the B2
(CsCl-like) structure under a highly external hydrostatic
pressure. Since Slater$^1$ first described this phase transition
in 1924, it has been considered to be one of the simplest
first-order transitions of alkali halides. However, its mechanism
still remains uncertain$^{14}$. If we restrict this transition to
a single-step process without intermediate structures$^{15}$,
there are some mechanisms that may be taken into account. One of
them is the Buerger mechanism$^{16}$, according to which pressure
contracts the B1 primitive rhombohedral cell along the [111] axis,
leading to the B2 phase. Another one is proposed as the WTM
mechanism$^{17}$ that concerns with the relative displacement of
consecutive [100] planes. The third mechanism$^{18}$ is closely
related to the Buerger one and concerns with the orientational
relations in a pressure-induced transition.

In our present work, we do not want to discuss the mechanism so we
just select the Buerger mechanism for its simplicity. We try to
explain the reason of transition using the energy minimization
technique. Here we introduce the Gibbs free energy G=U+PV, to
represent the total crystal energy, where U refers to the whole
cohesive energy (including the Coulomb part and the short-range
parts), P is the external pressure, V is the lattice volume, and
PV indicates the contribution of external pressure. We don't
consider the temperature factor, so that all the calculations are
under absolute zero.

Due to its too many freedom-degrees, the Gibbs free energy surface
is a hypersurface that can not be drawn on a 2-dimension plane.
However, we may fix the symmetry of the primitive cell as
$R\bar{3}m$ , change the lattice constant simultaneously with the
rhombohedral angle along the B1-B2 transition path, as described
in the Buerger mechanism. Then make the lattice constant as
X-coordinate, the rhombohedral angle as Y-coordinate and the Gibbs
free energy as the function of both lattice constant and
rhombohedral angle. Thus we can draw the Gibbs free energy surface
with a 2-dimension expression.

The Gibbs free energy surface at zero pressure is shown in Fig.2
(a). From the scheme we can see at zero pressure the B1 and
B2-structure are both local minima which mean stable and obviously
the B1-structure has a lower energy position than the
B2-structure. When we add an external pressure to the cell, the
whole energy surface will rise and the B1 minimum rises faster
relatively to the B2 minimum. But the B1 minimum remains lower
than the B2 minimum until the external pressure is high enough and
then a phase transition may occur. For this reason, we can expect
the B1-structure to be a more stable structure than the
B2-structure under a external pressure beneath the
transition-pressure.

In order to find the transition-point we calculate the Gibbs free
energies of relaxed B1 and B2-structure at different external
pressures by energy minimization. Both Gibbs free energies
increase with the external pressure and they have one crosspoint
at about 1.09GPa, beneath this value of pressure, the Gibbs free
energy of B1-structure is lower than that of B2-structure, while
above this value, the situation is just the opposite, as shown in
Fig.3. It indicates that, at an external pressure above 1.09GPa,
the B2-structure is more stable than B1, so the transition-point
is approximately 1.09GPa. This estimated transition-point accords
well with the experimental data about 10kbar in Ref.13.

From the Gibbs free energy surface at the transition pressure,
shown in Fig.2 (b), we can see the B1 and B2 minima are of the
same depth which means the B1 and B2-structure are of the same
stability at this time. However, it is difficult for a
B1-structure to turn into a B2-structure spontaneously by energy
minimization because there is still an energy barrier between the
two minima. With the further increase of external pressure the B1
minimum will become higher and higher. Finally, the B1-structure
will turn to a saddle point while the B2-structure remains a
minimum and there is clearly a downhill path from B1 to B2, as
shown in Fig.2 (c). Then we may achieve the process from B1-B2
using the energy minimization method.

\section{Molecular dynamic simulation of RbCl melting}
To perform our model to the thermal properties of B1-RbCl we
employ a MD technique. A detailed description of the molecular
dynamics method may be found elsewhere$^{19}$. In our work, the
simulations are performed using the MSI dynamics engine$^{20}$.
The successful runs rely on the size of system (number of
particles $N$), size of timestep ($\Delta t$), the total running
time (total steps of run $n_{steps}$), as well as the parameters
of the temperature and pressure control methods. After some test
runs we find the correct results can be obtained with $N=512$,
$\Delta t=5$ fs, $n_{steps}=10000$. And we choose Hoover method to
control the temperature and Andersen method to control the
pressure exactly at 0GPa. The NPT (constant N is the number of
particles, T is the temperature and P is pressure) ensemble is
adopted.

First we simulate the thermal expansion of B1-RbCl below the
melting point. The volume-temperature curve is shown in Fig.4. We
also approximately calculate the linear coefficient of thermal
expansion $\alpha$ at different temperatures, listed in Table III.
Compared with the experimental data$^{13}$, the calculated values
are reasonable.

However, we encounter a discrepance when we try to decide the
melt-point by heating the perfect lattice until it melts. The
abrupt change of volume which indicates the solid-liquid
transition occurs at the temperature of about 1260K, much higher
than the real melting temperature Tm 988K$^{13}$. And the lattice
remains solid at the temperature under which it is expected to
have been melted. This over-heat problem has been discussed in
some papers$^{21,22,23}$. It is said that the calculated
melt-point with a perfect lattice as the initial configuration
will be somewhat larger than the real melt-point, even larger by
an amount of the order of 20-30\%$^{21}$. And it is caused by the
lack of nucleation sites for the liquid phase as the crystal is
heated.

To avoid the over-heat problem and get the correct melt-point we
use a coexist-phase lattice as the initial
configuration$^{22,23}$, which means half of the particles are
solid and the rest are liquid. This coexist-phase lattice can be
obtained in the following manner: First, we build a superlattice
of 512 particles (256 Rb+ and 256 Cl- respectively). Then, with
half of the particles fixed, a MD run is carried out at a high
enough temperature to make sure the movable part can completely
melt. Thus the coexist-phase lattice with a common interface of
the solid part and the liquid part is obtained, as shown in Fig.6
(a). Using this coexist-phase lattice as the initial
configuration, a series of MD runs at different temperatures are
performed and the final configurations are saved and analyzed. We
find the abrupt change of volume happens at about 995K, very close
to the real melting temperature, as shown in Fig.5. From the saved
trajectory files we calculate the radial distribution function
(RDF) and the mean-square displacement (MSD) of the final
configurations at different temperatures. The abrupt changes of
RDF and MSD also indicate the occurrence of melting, as shown in
Fig.7, Fig.8. Further more, we pick up two final configurations
which represent complete-solidification and complete-melting as
shown in Fig.6 (b), (c). We can see the final configurations are
disparate on two sides of the melt-point even though they have the
same initial configurations. For comparison, the V-T curve with a
perfect lattice as the initial configuration is also shown in
Fig.5. The difference between the two calculated "melt-point" is
about 260K. So we prefer the coexist-phase method as the correct
method to calculate the melt-point of RbCl.

The MD simulations above are all carried out at 0 GPa. With an
external hydrostatic pressure adding to the superlattice, a more
serious over-heat problem will have to be taken into account.
Detailed discussion can be found in Ref.23.

Actually, the calculated "melt-point" with an initial
configuration of prefect lattice is not a denotation of melting,
but of the mechanical instability of the chosen model. One must be
sure not to confuse with these two conceptions.

\section{Conclusion}
Based on the M\"{o}bius inverse potential model, we have studied
the static structural and elastic properties of B1-RbCl crystal.
The calculated value are in good agreement with experimental data.
The high-pressure-induced B1-B2 transition of RbCl crystal is also
studied. The estimated transition-pressure is about 1.09GPa,
consistent with experimental data$^{13}$. Using the MD technique
we have simulated the melting of RbCl crystal at 0GPa and
calculated the correct melt-point with a coexist-phase method.

For these applications of the M\"{o}bius inverse potentials, we
consider it as a reliable model although it is derived just from
the cohesive energy curve without any empirical data employed.

However, there are still some defects in our model. First, using
the M\"{o}bius inversion method can only obtain the pairwise
potential, so that we must add a 3-body or more potential when it
is needed$^{24}$. Second, due to the infinite terms of summation
as , a quickly convergent E(x) is demanded. That is the reason why
we need to remove the Coulomb part from the total cohesive energy
while dealing with the ionic crystals.

Despite these flaws, we can see the simplicity and validity of
this M\"{o}bius lattice inversion method. This method has
successfully been used to obtain the interatomic potentials in
rare-earth metals and intermetallic compounds$^{25}$, and now the
interionic potentials of alkali halides.

\section*{Acknowledgements} Thanks for financial support from the
National Nature Science Foundation of China and the National
Advanced Materials Committee of China. Special support from the
973 Project in China, No.G2000067101, is hereby acknowledged.



\begin{figure}
\begin{center}
\includegraphics[scale=0.6]{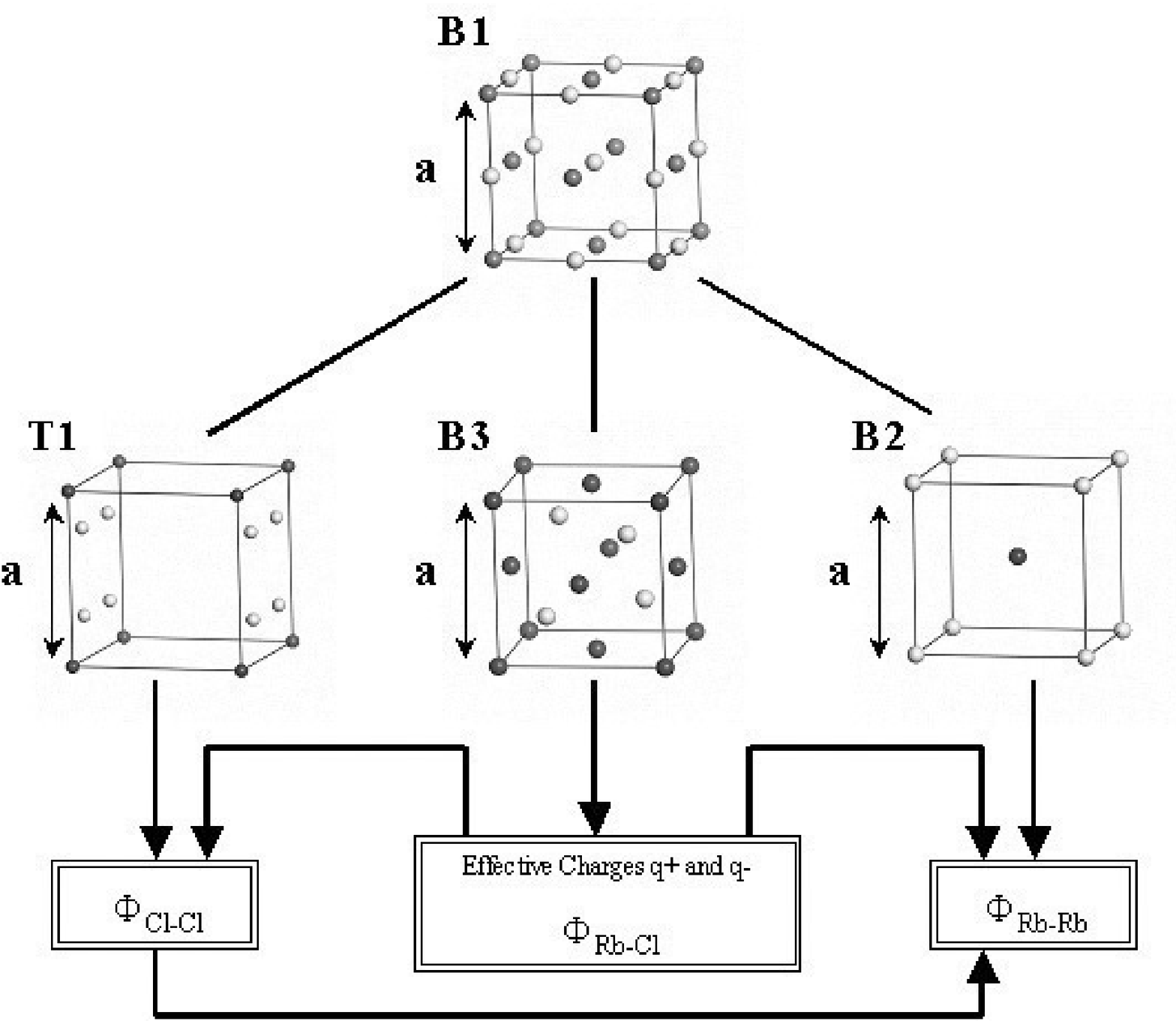}
\caption{The derivation of M\"{o}bius interionic potentials. Three
types of virtual structure are constructed: B3, T1, and B2, for
the purpose of extracting one from the totally three short-range
interactions: cation-anion, anion-anion, and cation-cation. Of
each energy curve, the long-range Coulomb energy has been
pre-subtracted from the total energy.}
\end{center}
\end{figure}

\begin{figure}
\begin{center}
\includegraphics[scale=1]{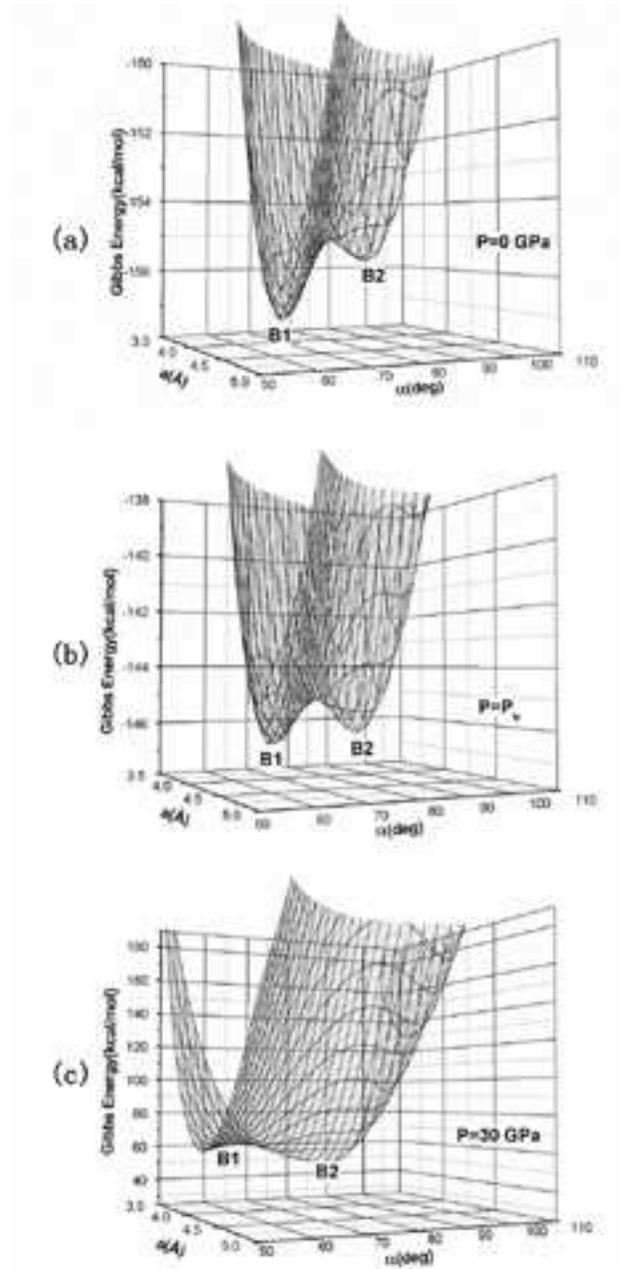}
\caption{The $0-K$ Gibbs free energy surfaces of R\={3}m symmetry
at (a) $zero$, (b) $P_{tr}$ and (c) $30GPa$, respectively.}
\end{center}
\end{figure}

\begin{figure}
\begin{center}
\includegraphics[scale=1]{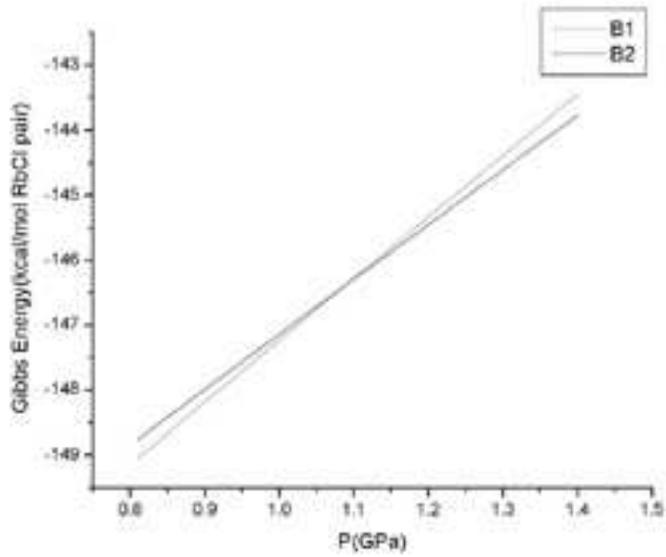}
\caption{The Gibbs free energies of relaxed B1 and B2-structure
RbCl crystals increase with external pressure. The two curves have
a crosspoint at $P=1.09GPa$. }
\end{center}
\end{figure}

\begin{figure}
\begin{center}
\includegraphics[scale=1]{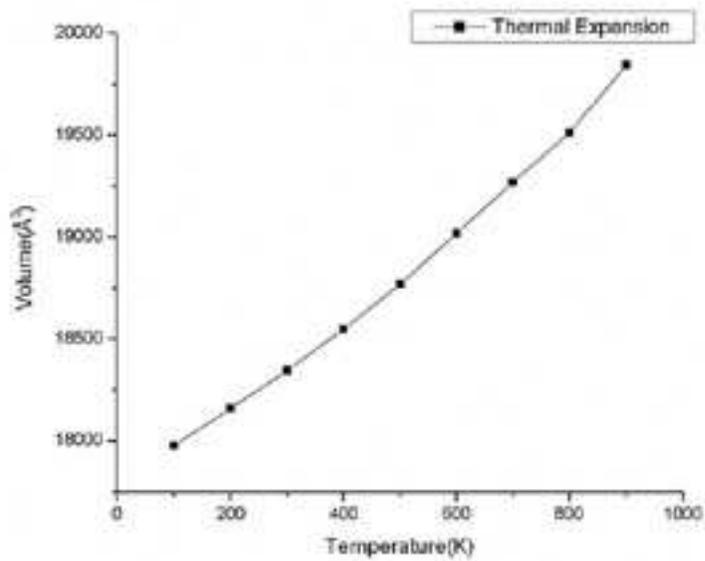}
\caption{The thermal expansion of perfect B1-RbCl crystal
calculated with molecular dynamics. Temperature varies from $100K$
to $900K$. The linear coefficient of thermal expansion is defined
as $(1/L)(dL/dT)$, where L is the lattice constant, and the
derivative is calculated as a central derivative with a
temperature step of $200K$.}
\end{center}
\end{figure}

\begin{figure}
\begin{center}
\includegraphics[scale=1]{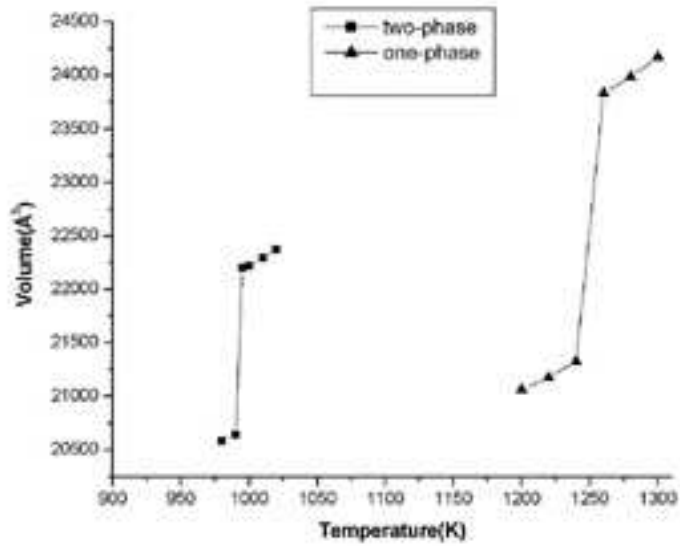}
\caption{The volume-temperature curves of the coexist-phase
lattice and the perfect lattice. The abrupt change of volume with
the one-phase method occurs at a temperature about $260K$ higher
than that of the two-phase method.}
\end{center}
\end{figure}

\begin{figure}
\begin{center}
\includegraphics[scale=0.6]{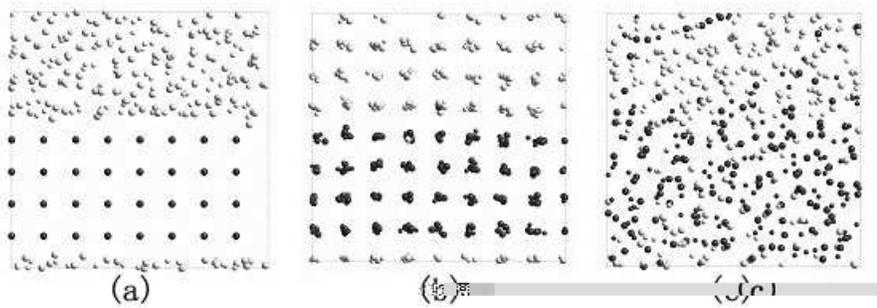}
\caption{The initial configuration of a coexist-phase lattice (a)
and possible final configurations (b)($980K$), (c)($1000K$) after
equilibrating the system at different temperatures. Dark circles
refer to ions that were initially in the solid phase; gray circles
refer to ions initially in the liquid phase. The system has almost
completely solidified as (b); and has almost completely melted as
(c).}
\end{center}
\end{figure}

\begin{figure}
\begin{center}
\includegraphics[scale=1.5]{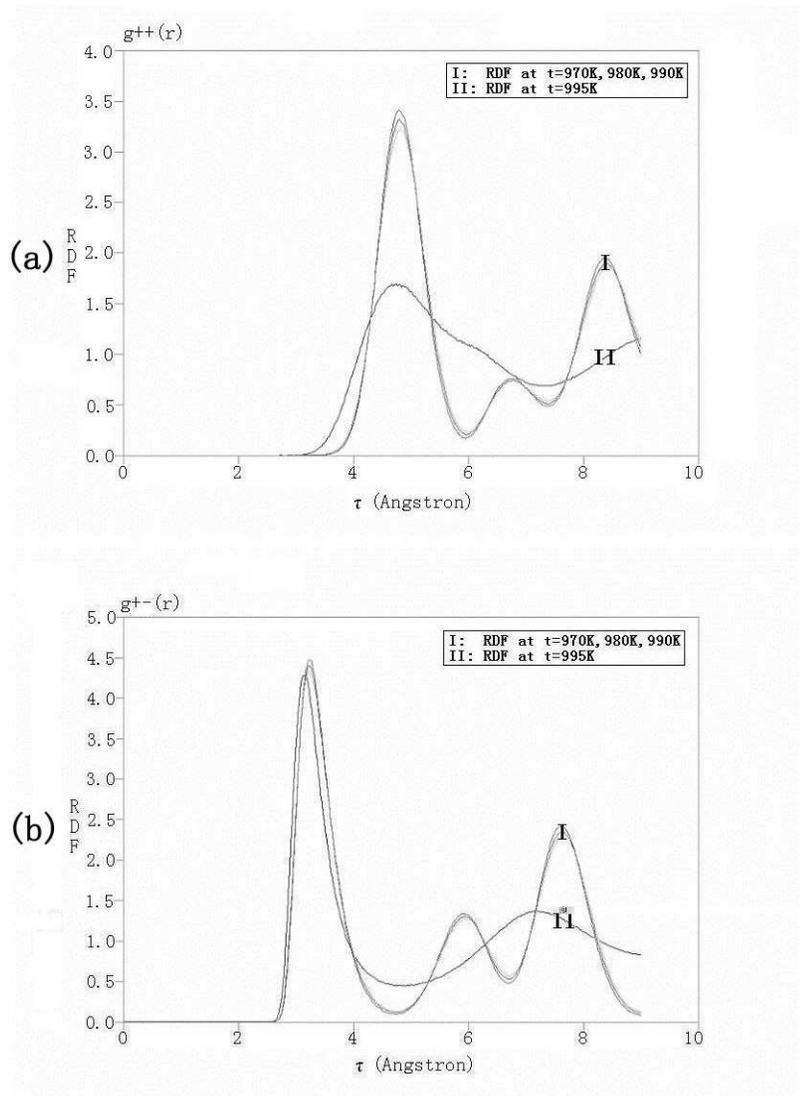}
\caption{The radial distribution function (RDF) of the final
configurations at different temperatures. The abrupt change of RDF
indicates the occurrence of melting. (a) shows the RDF of
cation-cation and (b) shows that of cation-anion.}
\end{center}
\end{figure}

\begin{figure}
\begin{center}
\includegraphics[scale=1.5]{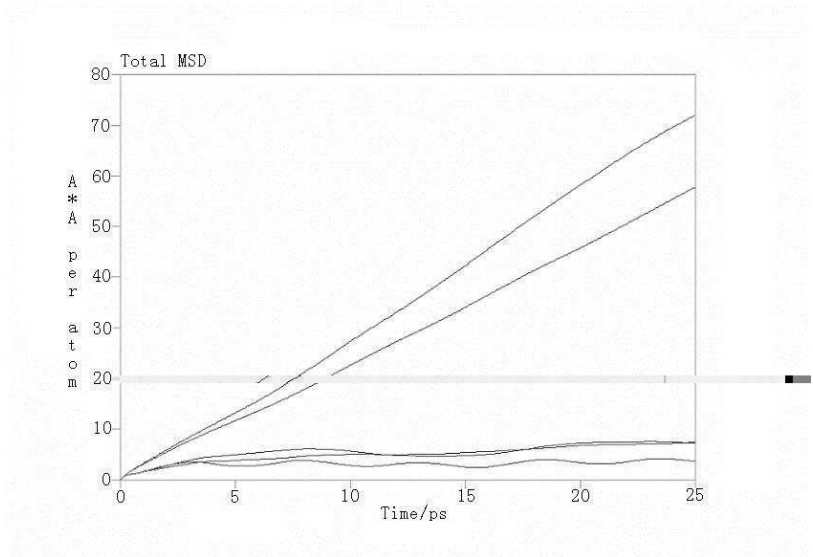}
\caption{The mean-square displacement (MSD) of all ions as the
function of time at different temperatures. The change of MSD
curves indicates the occurrence of melting: at low temperatures
the MSD oscillates near its balance, while it increases with time
at temperatures above melt-point.}
\end{center}
\end{figure}

\begin{figure}
\begin{center}
\includegraphics[scale=0.45]{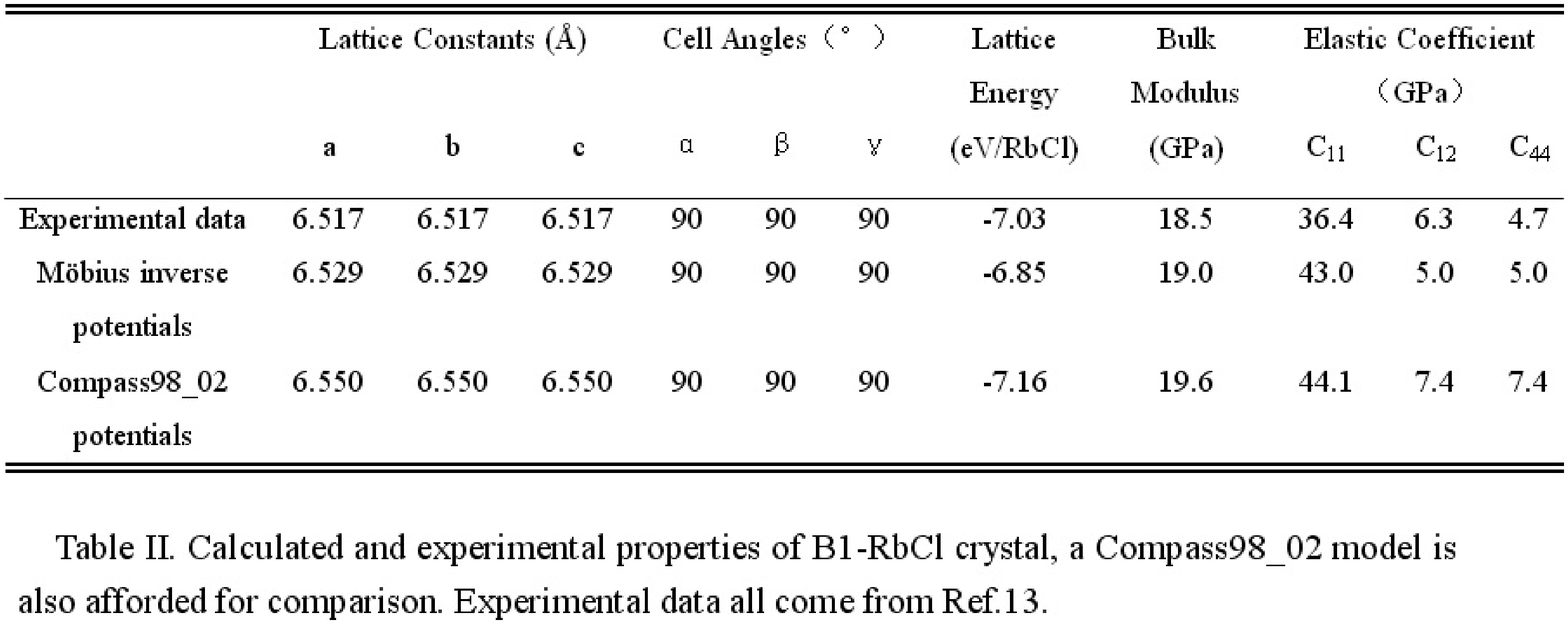}
\caption{Table II.}
\end{center}
\end{figure}

\begin{figure}
\begin{center}
\includegraphics[scale=0.5]{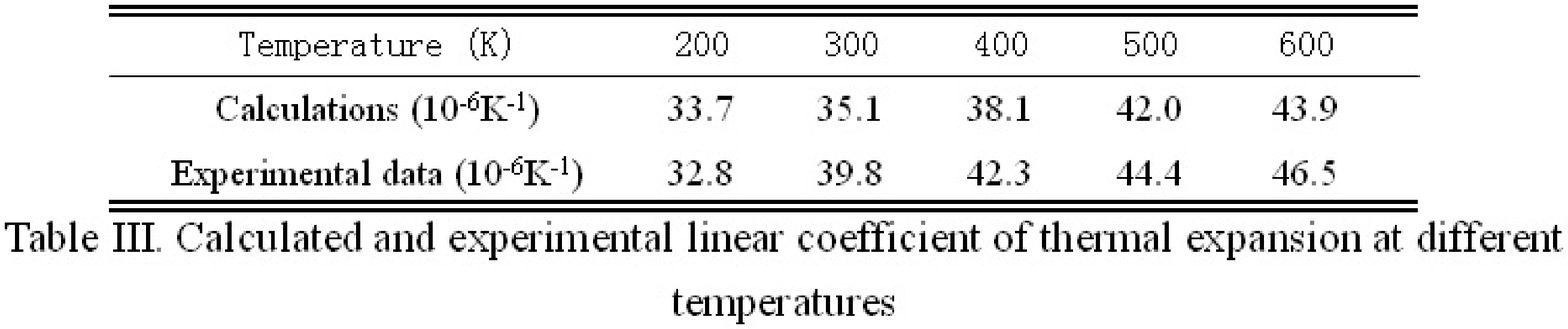}
\caption{Table III.}
\end{center}
\end{figure}

\end{document}